\documentclass{ifacconf}

\makeatletter
\let\old@ssect\@ssect 
\makeatother

\usepackage{algpseudocode}
\usepackage{algorithm}

\let\theoremstyle\relax 
\usepackage{amsmath,amsthm, amssymb}
\usepackage{graphicx}      
\usepackage{natbib}        
\usepackage{bm}
\usepackage{booktabs}
\usepackage{threeparttable}

\theoremstyle{plain}

\theoremstyle{definition}

\newtheorem{proposition}{Proposition}[section]

\newtheorem{remark}{Remark}[section]

\usepackage{xcolor, hyperref}

\definecolor{darkblue}{rgb}{0.0,0.0,0.6}
\hypersetup{colorlinks,breaklinks,linkcolor=darkblue,urlcolor=darkblue,anchorcolor=darkblue,citecolor=darkblue}
\makeatletter
\def\@ssect#1#2#3#4#5#6{%
\NR@gettitle{#6}
\old@ssect{#1}{#2}{#3}{#4}{#5}{#6}
}
\makeatother

\begin{document}

\begin{frontmatter}
	
\title{On Unified Adaptive Black-Litterman Mean-Variance Portfolio Management} 

\thanks[footnoteinfo]{This paper is partially supported by the Ministry of Science and Technology~(MOST), Taiwan, under Grant: MOST111--2813--C--007--021--H. This paper is a refined version of the earlier preprint \cite{li2023unified}, which is available on arXiv. $^{**}$Chung-Han Hsieh is the corresponding author.}

\author[First]{Chi-Lin Li} and 
\author[Second]{Chung-Han Hsieh}

\address[First]{Program of  Mathematical Finance and Financial Technology,\\
	Questrom School of Business, Boston University, USA.  \\
	(e-mail: \href{mailto:jadenli912@gmail.com}{jadenli912@gmail.com})}
\address[Second]{Department of Quantitative Finance, National Tsing Hua University, Hsinchu, 30004, Taiwan. \\
	(e-mail: \href{mailto:ch.hsieh@mx.nthu.edu.tw}{ch.hsieh@mx.nthu.edu.tw})}

\begin{abstract}
	This paper proposes a unified adaptive portfolio-management framework that combines factor-based view generation, Black-Litterman (BL) posterior estimation, EWMA covariance estimation, and mean-variance optimization. The key mechanism is a dynamic sliding window that adjusts the estimation horizon according to realized portfolio volatility, thereby updating factor estimates, BL posterior expected returns, and portfolio weights over time. In a ten-year empirical study of the top~100 market-capitalization constituents of the S\&P 500 with turnover transaction costs, the proposed method outperforms dynamic mean-variance optimization without BL views and provides stronger downside risk control, while its relative performance remains benchmark-dependent.
\end{abstract}

\begin{keyword}
	Control Applications, Portfolio Management, Black-Litterman Approach, Adaptive Optimization, Control-Oriented Finance
\end{keyword}
	
\end{frontmatter}

\section{Introduction}\label{section: introduction}
Portfolio optimization faces the fundamental challenge of balancing expected returns against estimation risk. While factor models provide systematic frameworks for return prediction, they often ignore investor-specific views. Conversely, the celebrated Black-Litterman (BL) approach, see \cite{black1992global}, incorporates views but typically relies on subjective inputs. This paper bridges these approaches by developing a unified adaptive BL framework that:
$(i)$ extracts data-driven views via factor models;
$(ii)$ incorporates these views through Black-Litterman posterior estimation;
$(iii)$ adapts to changing market conditions through volatility-responsive window sizing; and
$(iv)$ controls estimation risk through factor-model Elastic Net regularization.

The BL approach has been widely used and extended in portfolio construction; see, e.g., \cite{fabozzi2006incorporating, martellini2007extending}. However, its practical implementation still depends critically on how investor views are specified. Because BL views are typically formed subjectively, using analyst forecasts or institutional judgment, several studies have sought data-driven alternatives. Sentiment-based approaches require large linguistic datasets \cite{creamer2015can}; here, we instead use general factor models, as in \cite{kolm2020factor, spears2023view}, to generate BL views systematically.

 Expected-return estimation is also highly sensitive to sampling error \cite{luenberger2013investment}, and mean-variance optimization can amplify such errors into unstable portfolio weights; see, e.g., \cite{best1991sensitivity, britten1999sampling}. Motivated by regularized estimation methods \cite{chen2022black, spears2023view}, we use Elastic Net penalties in the factor-model estimation step, where the ridge component mitigates collinearity and the LASSO component shrinks weak factor loadings. The resulting factor-based views are then combined with a BL prior, and an EWMA update is used to estimate time-varying~covariance.

Finally, we make the estimation window adaptive. Dynamic-view BL methods have been studied in \cite{feng2016signal, guiso2018time, simos2021time}, while volatility-responsive sliding-window methods have been used in portfolio allocation and data-driven control \cite{hsieh2023asymptotic, hsieh2024solving, wang2022data}. Building on these ideas, our adaptive algorithm shortens the estimation window after increases in realized portfolio volatility and lengthens it in calmer periods. For a control-theoretic tutorial on stock-trading research, see \cite{barmish2024jump}.

%

\section{Unified Framework for Adaptive Portfolio Management} \label{section: preliminaries}
This section provides a unified framework for the adaptive portfolio management problem.

\subsection{Factor Model and Parameter Estimation} \label{subsection:FactorModelParams}
Consider $n$ risky assets. For $i=1,\dots,n$, 
let~$r_{i,\tau}$ denote the excess rate of return on asset $i$ (over the risk-free rate) at time~$\tau$, and~$\mathbf{F}_{\tau}$ be the vector of $J$ common factor returns at time~$\tau$. The general form for the factor model, see~\cite{tsay2005analysis}, is given by
\begin{align}\label{eq: factor models}
	r_{i,\tau} = \alpha_i + \bm{\beta}_i^\top \mathbf{F}_{\tau} + \varepsilon_{i,\tau}, \; i=1,\dots,n
\end{align}
where $\alpha_i$ is the asset-specific intercept,
$\mathbf{F}_{\tau} = [f_{1,\tau}, \dots, f_{J,\tau}]^\top$ is the vector of $J$
factor returns at time $\tau$ ($J<n$), $\bm{\beta}_i$ is the vector of factor
loadings for asset $i$, and $\varepsilon_{i,\tau}$ is the residual term.
We interpret $(\alpha_i,\bm{\beta}_i)$ as linear-projection coefficients, so the
residual is orthogonal to the regressors $[1,\mathbf F_\tau^\top]^\top$ in the usual~$L^2$ projection sense.

At each rebalancing time $t_k$, where $k=0, 1, 2, \dots$ indexes the number of rebalances,  the parameters $(\alpha_i, \bm{\beta}_i)$ for each asset $i$ are estimated using a rolling window of the $M_{t_k}$ most recent historical observations. Let these observations be indexed by $s=1, \dots, M_{t_k}$, corresponding to actual times $\tau_s \in \{t_k-M_{t_k}, \dots, t_k-1\}$. 
We denote the observed return of asset $i$ at the $s$-th point in this window as $r_{i,s}$ and the observed factor vector as~$\mathbf{F}_s$. The estimates $(\widehat{\alpha}_{i,t_k}, \widehat{\bm{\beta}}_{i,t_k})$ are obtained by solving the Elastic Net regression problem:
\begin{align}\label{problem: Elast Net regression for alpha and beta}
	\min_{\alpha_i, \bm{\beta}_i} \sum_{s=1}^{M_{t_k}} \left( r_{i,s} - (\alpha_i + \bm{\beta}_i^\top \mathbf{F}_s) \right)^2 + \lambda_2 \|\bm{\beta}_i\|_2^2 + \lambda_1 \|\bm{\beta}_i\|_1.
\end{align}
The regularization parameters are $\lambda_1 = \lambda \delta \ge 0$ and $\lambda_2 = \lambda(1-\delta) \ge 0$. Note that the penalty term $\|\bm{\beta}_i\|_2^2$ is the squared Euclidean norm.

\subsection{Factor-Based View Generation} \label{subsection: generation of views using factor model}
The estimated factor-model parameters $(\widehat{\alpha}_{i,t_k}, \widehat{\bm{\beta}}_{i,t_k})$ are used to generate \emph{investor views}. The view vector $\mathbf{q}_{t_k} \in \mathbb{R}^n$ at time $t_k$ has its $i$-th component defined as
$$
	q_{i,t_k}
	=
	\eta_{\alpha}\widehat{\alpha}_{i,t_k}
	+
	\widehat{\bm{\beta}}_{i,t_k}^{\top}
	\mathbf{F}_{\mathrm{view},t_k},
	\quad i=1,\dots,n,
$$
where $\eta_{\alpha}\in[0,1]$ controls the degree of intercept shrinkage. The case $\eta_{\alpha}=1$ gives the unshrunk factor-model view, whereas $\eta_{\alpha}=0$ uses only the factor-risk-premium component of the fitted model.

 The parameter $\eta_{\alpha}$ shrinks the fitted intercept because short-window
 intercept estimates are typically noisy. In particular, if the intercept is left
 unshrunk and the factor forecast is chosen as the mean of the same observations
 used to fit the regression, then the fitted view reduces to the rolling sample
 mean excess return. The shrinkage formulation therefore separates the systematic
 factor-premium component,
 $\widehat{\bm{\beta}}_{i,t_k}^{\top}\mathbf F_{\mathrm{view},t_k}$,
 from the asset-specific intercept component.

The vector $\mathbf{F}_{\mathrm{view},t_k}$ is a forecast of the factor vector for the upcoming period, constructed using information available up to $t_k-1$. In the empirical implementation, we use a maximum lookback length $L$ and estimate the expected factor return by averaging the available past factor observations up to this maximum. In particular, let~$\ell_{t_k}:=\min\{L,N_{t_k}\}$, where $N_{t_k}$ is the number of past factor observations available before $t_k$. Then
$$
\mathbf{F}_{\mathrm{view},t_k}
=
\widehat{\mathbb{E}}_{t_k-1}[\mathbf{F}_{t_k}]
=
\frac{1}{\ell_{t_k}}
\sum_{s=1}^{\ell_{t_k}}
\mathbf{F}_{t_k-s}.
$$

\subsection{Black-Litterman Posterior Estimation} \label{subsection: Black-Litterman Framework for Posterior Expected Returns}
The Black-Litterman (BL) approach combines a prior estimate of expected returns with investor views. Let~$\bm{\mu}_{t_k}$ be the true but unknown expected excess-return vector at time~$t_k$. The prior distribution for $\bm{\mu}_{t_k}$ is assumed to be Gaussian;~i.e.,
$$
	\bm{\mu}_{t_k} \sim \mathcal{N}(\bm{\Pi}_{t_k}, Q_{t_k}),
$$
where $\bm{\Pi}_{t_k}$ is the prior estimate of expected excess returns,\footnote{
	For example, from an equilibrium model like Capital Asset Pricing Model (CAPM): $\bm{\Pi}_{t_k} = \gamma_{t_k} \widehat{\Sigma}_{t_k} \mathbf{w}_{\mathrm{mkt}}$ for some scalar $\gamma_{t_k}$; see \cite{fabozzi2007robust, feng2016signal}.
	} and $Q_{t_k}$ is the covariance matrix representing confidence in this prior, typically $Q_{t_k} := \tau \widehat{\Sigma}_{t_k}$ for some scalar $\tau > 0$, where $\widehat{\Sigma}_{t_k}$ is the time-varying estimate of the asset excess-return covariance matrix, as will be defined in Section~\ref{subsection: covariance matrix estimation}.

The investor views $\mathbf{q}_{t_k}$, generated as in Section~\ref{subsection: generation of views using factor model}, are related to $\bm{\mu}_{t_k}$ via the linear model; see \cite{black1992global, he2002intuition}:
$$
\mathbf{q}_{t_k} = P \bm{\mu}_{t_k} + \bm{\varepsilon}_{q,t_k}, \quad \bm{\varepsilon}_{q,t_k} \sim \mathcal{N}(0, \Omega_{t_k}),
$$
where $P = I_{n \times n}$, since each component of $\mathbf q_{t_k}$ is an absolute view on the corresponding component of $\bm\mu_{t_k}$.  Here, $\Omega_{t_k}$ is the covariance matrix of the view errors. We specify $\Omega_{t_k}$ for parsimony as a diagonal matrix with entries $\kappa s_{i,t_k}^2$, where $s_{i,t_k}^2$ is the variance of the one-step-ahead forecast errors of the factor model for asset~$i$, estimated over the same rolling window of length $M_{t_k}$. In the implementation, these diagonal entries are bounded below by a small positive constant, ensuring~$\Omega_{t_k}\succ0$.

 We estimate the BL posterior mean through the weighted least-squares (WLS) form of the Gaussian prior-view model.
The posterior estimate, $\widehat{\bm{\mu}}_{t_k}$, is thus obtained by~solving:
 \begin{align} \label{eq:wLS}
		\min_{\bm{\mu}_{t_k}} \; \left\| V_{t_k}^{-1/2} \left( \begin{bmatrix} \bm{\Pi}_{t_k} \\ \mathbf{q}_{t_k} \end{bmatrix} - \begin{bmatrix} I \\ P \end{bmatrix} \bm{\mu}_{t_k} \right) \right\|_2^2
	\end{align}
where $V_{t_k} = \text{diag}(Q_{t_k}, \Omega_{t_k})$. When $V_{t_k}$ is singular or nearly singular, $V_{t_k}^{-1/2}$ is understood in the Moore--Penrose sense.

\subsection{EWMA Covariance Matrix Estimation} \label{subsection: covariance matrix estimation}
The time-varying covariance matrix of asset excess returns,~$\widehat{\Sigma}_{t_k}$, is estimated using an Exponentially Weighted Moving Average (EWMA) model. To make a decision at the start of the period beginning at $t_k$, we use information available up to and including time $t_k-1$. Let $\widehat{\Sigma}^{\mathrm{sam}}_{t_k}$ denote the sample covariance matrix computed from the asset excess returns in the current rolling window $[t_k-M_{t_k},t_k-1]$. The covariance matrix $\widehat{\Sigma}_{t_k}$ is updated as:
\begin{align}\label{eq:EWMA}
	\widehat{\Sigma}_{t_k}
	= \eta \widehat{\Sigma}_{t_{k-1}}
	+ (1-\eta)\widehat{\Sigma}^{\mathrm{sam}}_{t_k},
\end{align}
where $\eta \in [0,1]$ is the smoothing factor. The process is initialized with $\widehat{\Sigma}_{t_0}=\widehat{\Sigma}^{\mathrm{sam}}_{t_0}$ from the initial rolling window.
 This update is an EWMA-type smoothing of adaptive-window covariance estimates.

\subsection{Mean-Variance Portfolio Optimization}
Given the posterior expected excess-return estimate $\widehat{\bm{\mu}}_{t_k}$ and covariance estimate $\widehat{\Sigma}_{t_k}$ constructed above, the mean-variance allocation, e.g., \cite{markowitz1952portfolio}, at rebalancing time $t_k$ solves
\begin{align}\label{eq:mar}
	\max_{\mathbf{w} \in \mathcal{W}} \;
	\widehat{\bm{\mu}}_{t_k}^{\top}\mathbf{w}
	-
	\rho \mathbf{w}^{\top}\widehat{\Sigma}_{t_k}\mathbf{w},
\end{align}
where $\rho > 0$ is the risk-aversion coefficient, and
\[
\mathcal{W}
=
\left\{
\mathbf{w}\in\mathbb{R}^n:
\|\mathbf{w}\|_1\le 1,\;
|w_i|\le w_{\max}\ \text{for all }i
\right\}.
\]
The constraint $\|\mathbf{w}\|_1\le 1$ limits total gross exposure, while~$w_{\max}\in(0,1]$ is a single-asset position cap through the constraint $|w_i|\le w_{\max}$.
Here~$\mathbf{w}$ denotes the vector of risky-asset weights, and $\widehat{\bm\mu}_{t_k}$ is the posterior expected excess-return vector. Therefore, any residual allocation $1-\mathbf 1^\top\mathbf w$ is held in the risk-free asset.

\subsection{Adaptive Window Sizing}
A key adaptive component of the proposed framework is the data-window update rule, which adjusts the length of the historical window $M$ used for parameter estimation. This adaptation is performed at discrete rebalancing times, denoted by $t_k$ for $k=0, 1, 2, \dots$. The window size to be used at time $t_k$, denoted $M_{t_k}$, is determined by the realized volatility of the portfolio during the most recently concluded investment period. 

This approach is motivated by the concept of structural breaks in financial time series \cite{hansen2001new, tsay2005analysis}. In periods of high market turbulence, market dynamics shift rapidly, rendering older data less relevant for forecasting. A shorter data window is therefore preferable to make the model more responsive. Conversely, in stable market regimes, a longer data window provides more stable statistical estimates.

Let the investment period preceding the decision time~$t_k$ be the interval $[t_{k-1}, t_k - 1]$. Let $\sigma_{k-1}^{\text{realized}}$ denote the realized daily volatility of the portfolio $\mathbf{w}_{t_{k-1}}^*$ during this period. Let $\sigma_{\mathrm{ref}}$ denote the reference volatility used for comparison, initialized from the initial rolling window and updated after each rebalancing decision. The adaptive mechanism compares the most recent realized volatility~$\sigma_{k-1}^{\text{realized}}$ with the reference level $\sigma_{\mathrm{ref}}$. We define three market volatility regimes at time $t_k$:

\begin{itemize}
	\item \emph{Increasing Volatility Regime:} $\sigma_{k-1}^{\text{realized}} \geq (1+h) \sigma_{\mathrm{ref}}$. The window size is shortened to capture recent dynamics more effectively: $M_{t_k} = \max\{M_{\min},\lceil c_- M_{t_{k-1}} \rceil\}$.
	\item \emph{Decreasing Volatility Regime:} $\sigma_{k-1}^{\text{realized}} \leq (1 - h) \sigma_{\mathrm{ref}}$. The window size is lengthened to improve estimation stability: $M_{t_k} = \max\{M_{\min},\lceil c_+ M_{t_{k-1}} \rceil\}$.
	\item \emph{Stable Regime:} Otherwise. The window size remains unchanged: $M_{t_k} = M_{t_{k-1}}$.
\end{itemize}

Here, $h \in (0,1)$ is a sensitivity threshold, $c_- \in (0,1)$ is the shrinkage factor, $c_+ > 1$ is the expansion factor, and $M_{\min}$ is the minimum admissible window length. 
 The next rebalancing time is set by $t_{k+1}:=t_k+M_{t_k}$; hence, both the estimation window and trading frequency adapt to realized~volatility.

\begin{algorithm}[htbp]
	\scriptsize
	\caption{Adaptive BL--MV Algorithm}
	\label{algorithm: generating time-varying views}
	\begin{algorithmic}[1]
		\Require
		Portfolio dimension $n \geq 1$; number of factors $J$; minimum window size $M_{\min}\ge J+1$; initial window size $M_{t_0} \geq M_{\min}$;
		Elastic-Net parameters $(\lambda_1, \lambda_2)$;
		EWMA decay factor $\eta \in [0,1]$; volatility threshold~$h \in (0,1)$;
		view lookback length $L\geq 1$; intercept-shrinkage parameter $\eta_{\alpha}\in[0,1]$;
		risk-aversion $\rho>0$; window factors $c_{-}\!\in(0,1), c_{+}>1$.
		
		\Ensure
		Time-varying optimal weights $\{\mathbf{w}_{t_k}^*\}_{k=0}$ and portfolio characteristics.
		
		\State \textbf{Initialize:} Set rebalancing index $k \gets 0$, time $t_0$, initial window $M_{t_0}$.

		\State Compute $\widehat{\Sigma}^{\mathrm{sam}}_{t_0}$ from the initial rolling window $[t_0-M_{t_0},t_0-1]$ and set $\widehat{\Sigma}_{t_0}\gets \widehat{\Sigma}^{\mathrm{sam}}_{t_0}$.
		
		\State Using the initial window, obtain the initial weights~$\mathbf{w}_{t_0}^*$ by carrying out the factor-fit, view-generation, posterior-mean, and MV-optimization steps below.
		
		\State Initialize $\sigma_{\mathrm{ref}}$ as the realized daily volatility of $\mathbf{w}_{t_0}^*$ over the initial window $[t_0-M_{t_0},t_0-1]$.

		\While{trading horizon not reached}
		\State $k \gets k + 1$.
		\State Set current time $t_k \gets t_{k-1} + M_{t_{k-1}}$.
		\State Compute realized portfolio volatility $\sigma_{k-1}^{\text{realized}}$ over the interval $[t_{k-1}, t_k-1]$ using weights $\mathbf{w}_{t_{k-1}}^*$.
		
		\State \textbf{(Adaptive Logic)}\;
		\If{$\sigma_{k-1}^{\text{realized}} \ge (1+h)\,\sigma_{\mathrm{ref}}$}
		\State $M_{t_k} \gets \max\{M_{\min},\lceil c_{-}\,M_{t_{k-1}} \rceil\}$
		\ElsIf{$\sigma_{k-1}^{\text{realized}} \le (1-h)\,\sigma_{\mathrm{ref}}$}
		\State $M_{t_k} \gets \max\{M_{\min},\lceil c_{+}\,M_{t_{k-1}} \rceil\}$
		\Else
		\State $M_{t_k} \gets M_{t_{k-1}}$
		\EndIf
		\State $\sigma_{\mathrm{ref}} \gets \sigma_{k-1}^{\text{realized}}$.

	 \State Set $\ell_{t_k}\gets \min\{L,N_{t_k}\}$.

		\State \textbf{(Data Collection)}\; Collect asset excess returns and factor observations in the regression window
		$[t_k-M_{t_k},t_k-1]$, and collect factor observations in the view-estimation window
		$[t_k-\ell_{t_k},t_k-1]$.

		\State \textbf{(Factor Fit)}\; For each asset $i$, solve Equation~\eqref{problem: Elast Net regression for alpha and beta} using asset excess returns and factor observations from the regression window $[t_k-M_{t_k},t_k-1]$ to get $(\widehat{\alpha}_{i,t_k}, \widehat{\bm{\beta}}_{i,t_k})$.

		\State \textbf{(Views)}\; Generate view vector $\mathbf{q}_{t_k}$ using
		$\mathbf{F}_{\mathrm{view},t_k} = \frac{1}{\ell_{t_k}}\sum_{s=1}^{\ell_{t_k}}\mathbf{F}_{t_k-s}$
		and
		$q_{i,t_k}=\eta_{\alpha}\widehat{\alpha}_{i,t_k}
		+\widehat{\bm{\beta}}_{i,t_k}^{\top}\mathbf{F}_{\mathrm{view},t_k}$,
		for $i=1,\dots,n$.

		\State \textbf{(Covariance Update)}\; Compute $\widehat{\Sigma}^{\mathrm{sam}}_{t_k}$ from the current rolling window $[t_k-M_{t_k},t_k-1]$ and update $\widehat{\Sigma}_{t_k}$ using Equation~\eqref{eq:EWMA}.
		
		\State \textbf{(Posterior Mean)}\; Solve Equation~\eqref{eq:wLS} to obtain posterior mean~$\widehat{\bm{\mu}}_{t_k}$.

	\State \textbf{(Optimal Weights)}\; Solve the MV problem~\eqref{eq:mar} using $(\widehat{\bm{\mu}}_{t_k},\widehat{\Sigma}_{t_k})$ to find $\mathbf{w}_{t_k}^{\ast}$.

		\EndWhile
	\end{algorithmic}
\end{algorithm}

\medskip
\begin{proposition}[Causal Convex Implementability] \label{prop:causal-convex}
	For each rebalancing time $t_k$, suppose $M_{t_k}\ge 2$, $\ell_{t_k}\ge1$, $\eta\in[0,1]$, $\lambda_1,\lambda_2\geq 0$, $\rho>0$, $P=I_{n\times n}$, and $\Omega_{t_k}\succ0$. Then the following hold:
	\begin{enumerate}
		\item[$(i)$] $\widehat{\Sigma}_{t_k}\succeq 0$;
		\item[$(ii)$] the Elastic-Net regressions~\eqref{problem: Elast Net regression for alpha and beta}
		and the BL posterior problem~\eqref{eq:wLS} are convex; moreover, problem~\eqref{eq:wLS} is strictly convex, so the BL
		posterior mean $\widehat{\bm\mu}_{t_k}$ is~unique;
		\item[$(iii)$]  problem~\eqref{eq:mar} maximizes a concave objective over a non-empty, closed, bounded, convex set $\mathcal W$; hence its optimizer set is non-empty and compact;
		\item[$(iv)$] every quantity used to construct $\mathbf w_{t_k}^*$ is measurable with
		respect to the information available up to time~$t_k-1$.
	\end{enumerate}
\end{proposition}
\begin{proof}
	For $(i)$, the sample covariance matrix $\widehat\Sigma^{\mathrm{sam}}_{t_k}$ is positive semidefinite when $M_{t_k}\ge 2$. Since~\eqref{eq:EWMA} forms~$\widehat\Sigma_{t_k}$ as a convex combination of $\widehat\Sigma_{t_{k-1}}$ and $\widehat\Sigma^{\mathrm{sam}}_{t_k}$, positive semidefiniteness follows by induction.
	
	For $(ii)$, in the Elastic-Net regression problem~\eqref{problem: Elast Net regression for alpha and beta}, the squared-error loss is convex in $(\alpha_i,\bm\beta_i)$, while $\|\bm\beta_i\|_2^2$ and~$\|\bm\beta_i\|_1$ are convex penalties. Since the problem is unconstrained over the convex space $\mathbb R^{J+1}$, it forms a convex program. Similarly, 
for the BL posterior problem~\eqref{eq:wLS}, using~$P=I_{n\times n}$ and
$V_{t_k}=\operatorname{diag}(Q_{t_k},\Omega_{t_k})$, the objective is
\[
(\bm{\Pi}_{t_k}-\bm{\mu}_{t_k})^\top Q_{t_k}^{\dagger}
(\bm{\Pi}_{t_k}-\bm{\mu}_{t_k})
+
(\mathbf q_{t_k}-\bm{\mu}_{t_k})^\top \Omega_{t_k}^{-1}
(\mathbf q_{t_k}-\bm{\mu}_{t_k}),
\]
where $Q_{t_k}^{\dagger}$ denotes the Moore--Penrose inverse if $Q_{t_k}$ is singular.
Since $Q_{t_k}^{\dagger}\succeq0$, the first term is convex. Since~$\Omega_{t_k}\succ0$, the second term is strictly convex in $\bm{\mu}_{t_k}$. Hence the BL posterior problem is strictly convex and has a unique~minimizer.

	For $(iii)$, since $\widehat\Sigma_{t_k}\succeq 0$, the mapping
	$
	\mathbf w\mapsto
	\widehat{\bm\mu}_{t_k}^{\top}\mathbf w
	-\rho\mathbf w^\top\widehat\Sigma_{t_k}\mathbf w
	$
	is concave. The feasible set
	$
	\mathcal W=\{\mathbf w:\|\mathbf w\|_1\le 1,\ |w_i|\le w_{\max}\}
	$
	is non-empty, closed, bounded, and convex. By Weierstrass's extreme value theorem, an optimizer exists; concavity of the objective and convexity of $\mathcal W$ imply that the optimizer set is convex and compact.
	
	Finally, for $(iv)$, the regression window $[t_k-M_{t_k},t_k-1]$, the factor-view lookback, the EWMA covariance update, the CAPM prior inputs, and the adaptive-window comparison based on $\sigma_{k-1}^{\mathrm{realized}}$ all use only data available no later than~$t_k-1$. Hence the allocation $\mathbf w_{t_k}^*$ is causal.
\end{proof}

\subsection{Turnover Transaction Costs}
As demonstrated in the dynamic portfolio optimization literature by
\cite{brown2011dynamic, hautsch2019large, hsieh2023frequency}, transaction costs can significantly affect realized trading performance. In the empirical study, transaction costs are applied ex-post to portfolio turnover after each rebalancing decision.

Let $\nu\ge 0$ denote the proportional transaction-cost rate. At rebalancing time $t_k$, define turnover as
$
\text{Turnover}_{t_k} = \sum_{i=1}^n \left| w_{i,t_k}^{*} - w_{i,t_k}^{+} \right|,
$
where $w_{i,t_k}^{*}$ is the newly chosen target weight and $w_{i,t_k}^{+}$ is the drifted pre-trade portfolio weight immediately before rebalancing. The transaction cost deducted from wealth at~$t_k$ is
$
\text{Cost}_{t_k}
=
\nu\,\text{Turnover}_{t_k}\,V_{t_k}^{+},
$
where $V_{t_k}^{+}$ is the account value immediately before paying transaction costs. 
The initial portfolio formation at $t_0$ is treated as the starting allocation, so transaction costs are applied only to subsequent rebalancing trades. Turnover is computed over risky-asset weights only; no transaction cost is charged on the residual risk-free allocation.

\section{Empirical Studies: 100~Top-Capitalization S\&P 500 Portfolio} \label{section: illustrative examples}
We now evaluate our adaptive framework through extensive empirical studies using the top~100 market-capitalization constituents of the S\&P 500 at the beginning of the sample period and keep this universe fixed throughout the backtest. 
We first use daily closing prices for assets comprising the top 100 market cap assets of Standard and Poor's~500 (S\&P 500) constituents over ten years from January~1, 2013 to January~1, 2023.\footnote{
	The data is retrieved from CRSP and Compustat datasets, and access is authorized through the Wharton Research Data Service; see \cite{wrds2023}.
}

The sample contains both relatively stable market periods and the COVID-19 drawdown in the first half of~2020, providing a setting in which the volatility-responsive windowing mechanism can be evaluated.
We use a Four-Week U.S. Treasury bill\footnote{
	The data has been sourced from the~U.S. Department of the Treasury from $2013$ to $2023$.
} as the risk-free asset. 
The FF5 factor-model, BL view-generation, and mean-variance steps are applied to the 100 risky equity assets in excess-return space; any wealth not allocated to risky assets is held in the Treasury bill.

To evaluate the trading performance, we use the following metrics.
The realized portfolio excess return is computed as the portfolio return net of the risk-free return. We denote its annualized sample mean by~$\overline{r^p}$, its annualized volatility by~$\sigma$, and its annualized Sharpe ratio by~\texttt{SR}.
To measure downside risk, we let $d^*$ denote the \emph{maximum percentage drawdown} of the realized wealth path and report the Calmar ratio, $\texttt{CR}:=\overline{r^p}/d^*$, which measures annualized excess return per unit of maximum drawdown. We refer the reader to \cite{hsieh2017inefficiency, hsieh2017drawdown} for further details on drawdown-based control-theoretic portfolio analysis.

\subsection{FF5 Factor Model} \label{section: Factor Model}
We implement our framework using the Fama-French five-factor model (FF5), a widely used benchmark model in empirical asset-pricing research; see \cite{fama2015five}. The FF5 model captures market, size, value, profitability, and investment factors, providing a parsimonious representation of systematic risk for our portfolio universe.

In our implementation, we derive the $\alpha_i$ and~$\bm{\beta}_i$ parameters used for view generation by estimating the following time-series regression for each risky asset $i$ using the historical data within each sliding window:
\begin{align*}
	r_{it} - r_{ft} 
	&= \alpha_i + \beta_{i, \texttt{Mkt}}(r_{mt}-r_{ft}) + \beta_{i,\texttt{SMB}}\texttt{SMB}_t + \beta_{i,\texttt{HML}}\texttt{HML}_t \\
	&\quad + \beta_{i,\texttt{RMW}}\texttt{RMW}_t + \beta_{i,\texttt{CMA}}\texttt{CMA}_t + \varepsilon_{it}
\end{align*}
where $r_{it}$ is the total return of asset $i$ at time $t$, $r_{ft}$ is the risk-free rate, $(r_{mt}-r_{ft})$ is the market excess return factor, and $\texttt{SMB}_t, \texttt{HML}_t, \texttt{RMW}_t, \texttt{CMA}_t$ are the returns of the other Fama-French factors at time $t$, which can be obtained from Kenneth French Data Library, see~\cite{frenchdatalibrary}. 
The estimated intercept $\widehat{\alpha}_i$ and factor loadings
$\widehat{\bm{\beta}}_i =
[\widehat{\beta}_{i,\texttt{Mkt}},\widehat{\beta}_{i,\texttt{SMB}},\dots,
\widehat{\beta}_{i,\texttt{CMA}}]^\top$
are combined with the factor-premium estimate $\mathbf F_{\mathrm{view},t}$ to form
the excess-return~view
\[
q_{i,t}
=
\eta_{\alpha}\widehat{\alpha}_{i,t}
+
\widehat{\bm{\beta}}_{i,t}^{\top}\mathbf F_{\mathrm{view},t}.
\]
In the reported empirical specification, $\eta_{\alpha}=0$, so the view is the pure
factor-premium view
$
q_{i,t}
=
\widehat{\bm{\beta}}_{i,t}^{\top}\mathbf F_{\mathrm{view},t}.
$

 \subsection{Out-of-Sample Trading Performance}
 \label{section: Trading Performance}
 Using an initial account with \$1,000,000 and initial window size $M_{t_0}=50$ trading days, we carry out the adaptive~BL--MV Algorithm~\ref{algorithm: generating time-varying views} with volatility threshold~$h:=0.1$, shrinkage factor $c_- = 0.8$, and expansion factor~$c_+ = 1.25$.

 For the reported empirical specification, we set the CAPM-prior risk-aversion parameter to $\gamma=10$, the MV risk-aversion parameter to $\rho=2.5$, the intercept-shrinkage parameter to $\eta_{\alpha}=0$, the minimum window length to~$M_{\min}=7$, and the maximum factor-view lookback length to (approximately three-year) $L=756$ trading days.\footnote{ 
 	The bounded sensitivity exercise considered $\gamma \in \{1,2.5,5,10\}$, $\rho \in \{0.5,1,2.5,5\}$, $\eta_{\alpha} \in \{0,0.1, 0.5, 1\}$, and $L \in \{756, 1008, 1260\}$. When fewer than $L$ past factor observations are available, the factor-view estimate uses all available past observations. The position cap is set to $w_{\max}=0.10$.  
 }
 After initialization, the window size is updated adaptively at each rebalancing time:
 if the current realized portfolio volatility~$\sigma$ satisfies
 $\sigma \geq (1+h)\sigma_{\mathrm{ref}}$, we set
 $M \leftarrow \max\{M_{\min},\lceil c_-M\rceil\}$; if
 $\sigma \leq (1-h)\sigma_{\mathrm{ref}}$, we set
 $M \leftarrow \max\{M_{\min},\lceil c_+M\rceil\}$; otherwise, $M$ is maintained.

 In covariance matrix estimation, we use $\eta = 0.2$ in Equation~\eqref{eq:EWMA}, so the current rolling-window sample covariance receives weight $1-\eta=0.8$. 
 The shrinkage factor $c_- = 0.8$ controls how aggressively the window is shortened after a volatility increase, whereas the expansion factor $c_+ = 1.25$ controls how quickly the window is lengthened after a volatility decrease.

\begin{figure}[htbp]
	\centering 
	\includegraphics[width=1.0\linewidth]{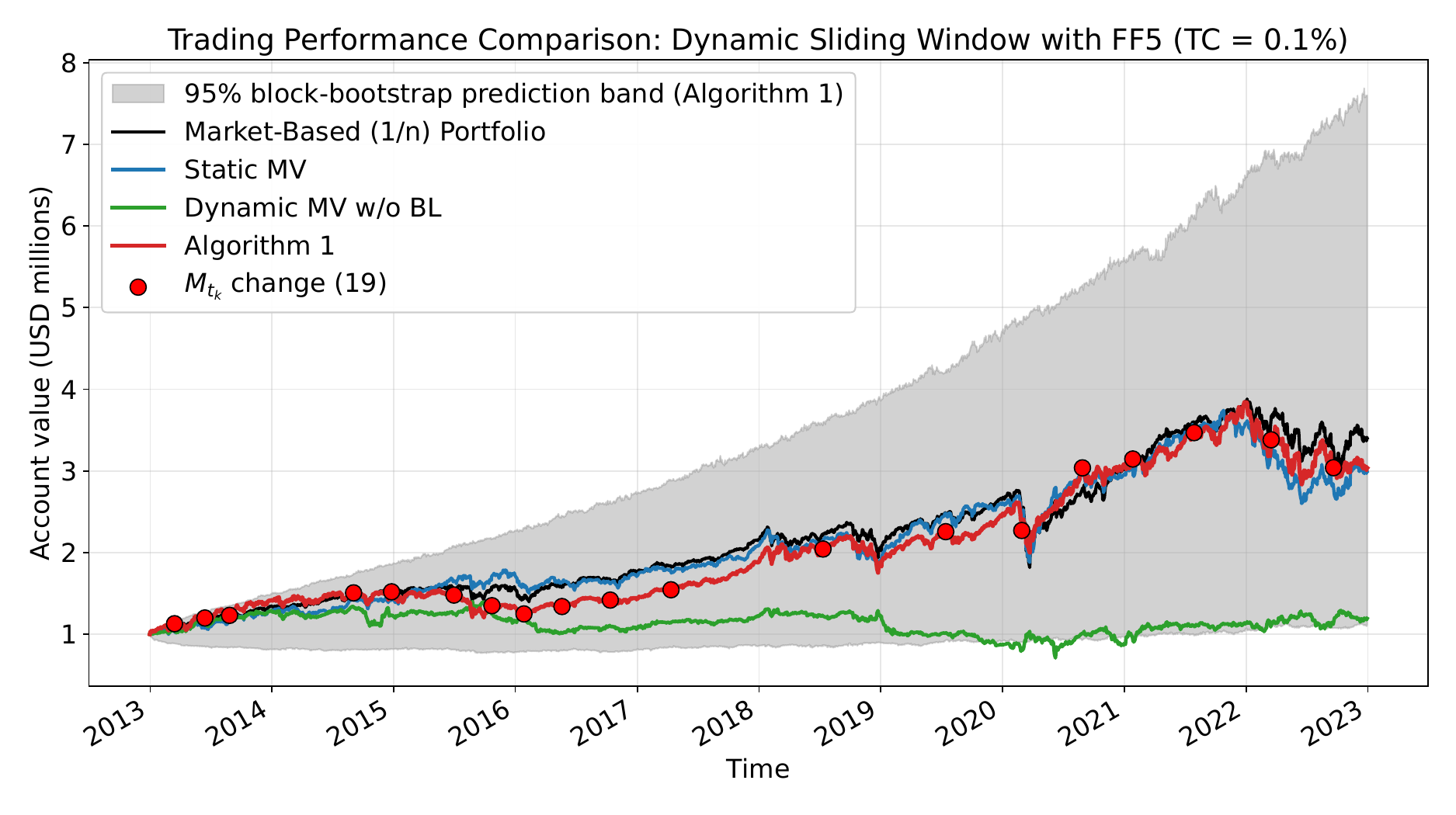}
	\caption{Out-of-Sample Trading Performance under FF5 Views with $\mathrm{TC}=0.1\%$.}
	\label{fig: Dynamic Sliding Window Size within Fama-French Five-Factor Model (2013-2023)}
\end{figure}

Figure~\ref{fig: Dynamic Sliding Window Size within Fama-French Five-Factor Model (2013-2023)} shows the portfolio account value trajectories under transaction cost $\mathrm{TC}=0.1\%$ for the market-based portfolio (\texttt{Market-Based Portfolio}),\footnote{
	The market-based $1/n$ portfolio represents an equally weighted portfolio over the 100 risky assets, i.e.,~$w_i=\frac{1}{100}$ for all $i=1, 2, \dots, n=100$, with zero residual risk-free allocation.
	} static mean-variance portfolio without BL model (\texttt{Static MV}), dynamic mean-variance portfolio without BL model (\texttt{Dynamic MV w/o BL}), and the dynamic MV portfolio with dynamic BL (\texttt{Algorithm 1}), which is generated by the adaptive BL--MV  Algorithm~\ref{algorithm: generating time-varying views}. 
The gray-shaded region reports a 95\% moving-block bootstrap prediction band for the Algorithm~\ref{algorithm: generating time-varying views} account-value trajectory. 
Specifically, we compute the realized daily returns implied by the Algorithm~\ref{algorithm: generating time-varying views} wealth path, resample these returns in overlapping blocks, construct 1,000 bootstrapped wealth trajectories, and report the 2.5th and 97.5th percentiles at each date. 
The block bootstrap is used only to visualize path uncertainty and does not affect the realized performance metrics reported in Table~\ref{table:Different Transaction Costs (FF5)}. 
In the figure, the red dots indicate the instances when the window size~$M_{t_k}$ is adjusted.

\begin{remark}[Computational Implementation]
	In the empirical implementation, the convex subproblems in
	Algorithm~\ref{algorithm: generating time-varying views} can be solved using
	CVXPY; see \cite{diamond2016cvxpy}. A typical full rebalancing step takes under~$10$ seconds on a typical~$3.50$ GHz laptop with $16$ GB RAM.
\end{remark}

 \begin{table}[htbp]
 	\centering
 	\scriptsize
 	\setlength{\tabcolsep}{4pt}        
 	\caption{Trading Performance (2013--2022): FF5 with Different Transaction Costs}
 	\label{table:Different Transaction Costs (FF5)}
 	\begin{threeparttable}
 		\begin{tabular}{l rrrrr}
 			\toprule
 			Portfolio Strategy & $\overline{r^p}$ (\%) & $\sigma$ (\%) & \texttt{SR} & $d^*$ (\%) & \texttt{CR} \\
 			\midrule
 			 \multicolumn{6}{l}{\emph{TC = 0\%}} \\
 			Dynamic MV w$/$o BL &  4.12 & 18.33 & 0.22 & 45.22 & 0.09 \\
 			Algorithm 1         & 12.49 & 18.25 & 0.68 & \textbf{26.34} & \textbf{0.47} \\
 			\midrule
 			\multicolumn{6}{l}{\emph{TC = 0.01\%}} \\
			Dynamic MV w$/$o BL &  4.03 & 18.33 & 0.22 & 45.48 & 0.09 \\
			Algorithm 1         & 12.46 & 18.25 & 0.68 & \textbf{26.35} & \textbf{0.47} \\
 			\midrule
 			 \multicolumn{6}{l}{\emph{TC = 0.1\%}} \\
			Dynamic MV w$/$o BL &  3.17 & 18.34 & 0.17 & 47.74 & 0.07 \\
 			Algorithm 1         & 12.24 & 18.25 & 0.67 & \textbf{26.41} & \textbf{0.46} \\
 			\midrule
 			\multicolumn{6}{l}{\emph{TC = 1\%}} \\
 			Dynamic MV w$/$o BL & $-5.41$ & 18.82 & $-0.29$ & 68.16 & $-0.08$ \\
 			Algorithm 1         &    10.05 & 18.33 &    0.55 & \textbf{27.01} & \textbf{0.37} \\
 			\midrule
 			\multicolumn{6}{l}{\emph{Benchmark (Buy-and-Hold)}} \\
 			1$/$n Portfolio & {13.09} & {17.11} & {0.76} & 34.36 & 0.38 \\
 			Static MV       & 11.91 & 17.87 & 0.67 & 30.39 & 0.39 \\
 			\bottomrule
 		\end{tabular}
 		\begin{tablenotes}
 			\scriptsize
 			\item Note: $\overline{r^p}$ = annualized mean excess return; $\sigma$ = annualized volatility;
 			$\texttt{SR}$ = annualized Sharpe ratio; $d^*$ = maximum percentage drawdown;
 			$\texttt{CR}$ = Calmar ratio, $\overline{r^p}/d^*$. 
 			The 1$/$n Portfolio and Static MV strategies are reported once because they are buy-and-hold benchmarks and do not vary with the transaction-cost level. 
 			Boldface highlights the downside-risk entries, $d^*$ and \texttt{CR}, for Algorithm~1 among the transaction-cost-sensitive strategies at each transaction-cost level.
 		\end{tablenotes}
 	\end{threeparttable}
 \end{table}

Table~\ref{table:Different Transaction Costs (FF5)} shows that Algorithm~\ref{algorithm: generating time-varying views}
substantially improves over the dynamic MV strategy without BL views across all transaction-cost levels.
It also delivers the smallest maximum drawdown among the transaction-cost-sensitive strategies.
However, the equal-weighted benchmark retains the highest Sharpe ratio and annualized excess return
in this sample, consistent with the empirical robustness of $1/n$ documented by \cite{demiguel2009optimal}. Thus the empirical advantage of the proposed method is strongest in downside-risk
control and relative improvement over dynamic MV without BL, rather than uniform dominance over all
benchmarks.

\subsection{Stress Test under Factor-Model Misspecification}
\label{section: Monte-Carlo Based Robustness Test}

The purpose of this section is to test whether the proposed factor-view mechanism creates artificial gains when the data-generating process contains no FF5 factor structure. We therefore simulate asset returns from a correlated geometric Brownian motion (GBM) model calibrated from the same empirical sample used in Section~\ref{section: Trading Performance}. Specifically, for each Monte-Carlo path we generate daily simple returns~from
$$
\log(1+\mathbf R_t^{\mathrm{GBM}})
=
\widehat{\mathbf m}
+
\widehat L \mathbf z_t,
\qquad
\mathbf z_t\sim\mathcal N(0,I),
$$
where $\widehat{\mathbf m}$ is the empirical mean vector of daily log returns and $\widehat L\widehat L^\top$ is the empirical covariance matrix of daily log returns. Thus the simulated returns preserve the historical cross-sectional covariance structure, but are not generated from the FF5 factors.

During this simulation, Algorithm~\ref{algorithm: generating time-varying views} is run without modification. It continues to estimate the FF5 factor model and construct views using historical FF5 factor observations available before each rebalancing date. Hence the factor-view model is intentionally misspecified in this experiment.

\begin{table}[htbp]
	\centering
	\scriptsize
	\caption{GBM Misspecification Stress Test: Correlated GBM with TC=0\%}
	\label{table: Robustness Test Via Monte-Carlo Simulations FF5}
	\begin{tabular}{l rrrrr}
		\toprule
		Portfolio Strategy & $\overline{r^p}$ (\%) & $\sigma$ (\%) & \texttt{SR} & $d^*$ (\%) & \texttt{CR} \\
		\midrule
		1$/$n Portfolio       & 15.39 & 17.85 & 0.84 & 27.91 & 0.52 \\
		Static MV             & 13.15 & 19.85 & 0.60 & 33.17 & 0.38\\
		Dynamic MV w$/$o BL   &  4.06 & 18.54 & 0.23 & 46.92 & 0.10 \\
		Algorithm~\ref{algorithm: generating time-varying views}
		&  7.97 & \textbf{12.29} & 0.66 & \textbf{22.08} & 0.37 \\
		\bottomrule
	\end{tabular}
\end{table}

Table~\ref{table: Robustness Test Via Monte-Carlo Simulations FF5} reports median performance across 100 Monte-Carlo paths. As expected, Algorithm~\ref{algorithm: generating time-varying views} does not dominate the equal-weighted benchmark in Sharpe ratio when the FF5 view model is misspecified. However, it substantially reduces downside risk relative to the optimized MV benchmarks.

\medskip
 \begin{remark}[Regularization Units] \rm
 	The reported factor-regression penalties $\lambda_1=\lambda_2=0.5$  are in percentage-return units; with decimal returns and factors, the equivalent penalties are $\lambda_1=\lambda_2=0.5/100^2$.
 \end{remark}

\section{Concluding Remarks} \label{section: concluding remarks}
This paper presents a unified adaptive Black-Litterman mean-variance framework that combines factor-based view generation, Elastic-Net factor estimation, EWMA covariance estimation, and volatility-responsive window sizing. The resulting algorithm updates factor estimates, BL posterior expected returns, covariance estimates, and portfolio weights using only information available before each rebalancing date.

Proposition~\ref{prop:causal-convex} establishes causal implementability and convexity of the optimization subproblems. In the empirical study, Algorithm~\ref{algorithm: generating time-varying views} substantially outperforms dynamic mean-variance optimization without BL views and achieves the strongest downside-risk performance among the transaction-cost-sensitive strategies. Its relative Sharpe ratio and return performance, however, remain benchmark-dependent.

\small
\bibliography{refs}

\end{document}